\documentclass{article}
\usepackage{authblk}
\usepackage[font=small,labelfont=bf,justification=justified]{caption}
\usepackage{float}
\usepackage{mathpazo}
\usepackage{graphicx}
\usepackage[margin=0.7in]{geometry}
\date{}

\title{\textbf{Directional Second Harmonic Generation Controlled by Sub-wavelength Facets of an Organic Mesowire}}
\begin{document}

\author[1]{Deepak K. Sharma}
\author[1]{Shailendra K. Chaubey}
\author[1]{Adarsh B. Vasista}
\author[1]{Jesil Jose}
\author[1,2]{Ravi P N Tripathi}
\author[3]{Alexandre Bouhelier}
\author[1,4,*]{G V Pavan Kumar}

\affil[1]{Photonics and Optical Nanoscopy Laboratory, Department of Physics, Indian Institute of Science Education and Research (IISER), Pune-411008, India}
\affil[2]{Current Address: Jai Prakash Vishwavidyalaya, Chapra 841301, India}
\affil[3]{Laboratoire Interdisciplinaire Carnot de Bourgogne, UMR 6303 CNRS,Universit\'e de Bourgogne Franche-Comt\'e, 21000 Dijon, France}
\affil[4]{Center for Energy Sciences, Indian Institute of Science Education and Research (IISER), Pune-411008, India}

\affil[*]{Corresponding author: pavan@iiserpune.ac.in}



\maketitle
\begin{abstract}
Directional harmonic generation is an important property characterizing the ability of nonlinear optical antennas to diffuse the signal in well-defined region of space. Herein, we show how sub-wavelength facets of an organic molecular mesowire crystal can be utilized to systematically vary the directionality of second harmonic generation (SHG) in the forward scattering geometry. We demonstrate this capability on crystalline diamonoanthraquinone (DAAQ) mesowires with subwavelength facets. We observed that the radial angles of the SHG emission can be tuned over a range of 130 degrees. This angular variation arises due to spatially distributed nonlinear dipoles in the focal volume of the excitation as well as the geometrical cross-section and facet orientation of the mesowire. Numerical simulations of the near-field excitation profile corroborate the role of the mesowire geometry in localizing the electric field. In addition to directional SHG from the mesowire, we experimentally observe optical waveguiding of the nonlinear two-photon excited fluorescence (TPEF). Interestingly, we observed that for a given pump excitation, the TPEF signal is isotropic and delocalized, whereas the SHG emission is directional and localized at the location of excitation. All the observed effects have direct implications not only in active nonlinear optical antennas, but also in nonlinear signal processing.\\
\textit{Accepted in Applied Optics:
“\copyright 2018 Optical Society of America. One print or electronic copy may be made for personal use only. Systematic reproduction
and distribution, duplication of any material in this paper for a fee or for commercial purposes, or modifications of the content
of this paper are prohibited” }
\end{abstract}


\section{Introduction}

 Localized manipulation and directional, linear and nonlinear emission of light at sub-wavelength scale is one of the important goals of nanophotonics. To this end, a variety of nanostructures made of plasmonic\cite{ojant,smcr,nn,brongnm,hhbook,npz,teri,ojnano,alex,kth,ogatanre,ogataapl,ogataome} and dielectric objects\cite{maglight,luksci,kivopn,kivtica,jopt,smge,marin1,Mokkapati} have been realized, thanks to advancement in nanofabrication methods including top-down and bottom-up approaches.\\
     In recent times, nanophotonic operations of dielectric and semiconducting nanostructures have emerged as an exciting prospect in photonics\cite{kivopn,nb2,nb1,ritesh,kruk}. Given that ohmic loses in dielectric nanostructures\cite{luksci,jopt} are mitigated compared to plasmonic metals, they can be utilized in applications where Joule heating is detrimental. Certain dielectric and semiconducting nanostructures with high permittivity also facilitate multipolar Mie resonances \cite{kivtica,nb2} that can be harnessed for developing nano-optical biosensing\cite{kivbio,nb1}, novel optical antennas\cite{kivmag,kivtica}, and optimizing nonlinear frequency conversion\cite{kivshg,nb1}. As inherent losses are minimized in such nanostructures, they can sustain greater excitation powers, and hence provide an excellent opportunity to improve nonlinear optical processes such as second harmonic generation (SHG), third harmonic generation (THG), ultrafast optical switching, and many more \cite{luksci,kivopn,PhysRevB,Osgood09,jerry}.  \\
    An emerging prospect\cite{luksci,kivopn,kivtica,jopt} in dielectric nanophotonics is to combine the unique properties of optical antennas with nonlinear optics. In this direction, interesting developments have occurred recently, such as: the observation of directional SHG and THG emission from silicon nanostructures \cite{kivshg, kivthg, kivprb}; efficient THG from germanium nano-disc \cite{smge}; nonlinear vector beam generation\cite{vector}, SHG radiation pattern engineering \cite{apshape}, and polarization control\cite{polshg} in Aluminum Gallium Arsenide (AlGaAs) nanoantenna; metal-dielectric hybrid nonlinear nano-antennas\cite{nnsm, hg1, hg2,hg3,lsahybrid,scirep,ren} and metasurfaces \cite{haykap} with a specific directional emission. \\
A majority of the studies on dielectric nonlinear nanophotonics have utilized inorganic nanostructures such as silicon, germanium and AlGaAs. The general approach to prepare such structures is based on top-down nanofabrication either using high-power ultrafast lasers \cite{maglight,ncmag}  or electron beam lithography\cite{vector,haykap}. Such top-down approaches generally result in polycrystalline nanomaterials, which can affect the yield of nonlinear optical processes.
As a complementary approach, organic nanostructures can be prepared in single crystalline form using vapour phase methods, and can be potentially utilized for nonlinear optical antenna applications. Furthermore, organic molecular nanostructures have some advantages \cite{nporganic, C4C,zhao2014,chandrasekar,venk}  such as low cost processing, flexibility of deposition on desired substrate, ultrafast and large nonlinear response, broadband spectral tunability by tailoring the structure of molecules and their arrangement in crystals.\\
    In the context of nonlinear optical antennas based on dielectric nanostructures, there are two important aspects that have to be considered\cite{apshape}: first is the efficiency of nonlinear optical frequency conversion, and the second is the directionality of the nonlinear radiation generated by the dielectric nanostructures. Over the years, a variety of organic crystals have been harnessed for efficient nonlinear optical frequency conversion, and crystalline organic nanostructures have been explored in the context of their nonlinear microscopy \cite{z2prl,brasaop,Chandra2017}. However a relatively unexplored aspect of organic nanostructure is their utility as nonlinear optical antennas, especially their directional nonlinear emission characteristics. \\
   In this paper, we show, how sub-wavelength facets of an organic molecular mesowire crystal made of diaminoanthraquinone (DAAQ) can be harnessed to actively control forward scattered SHG pattern. We test the ability of the mesowires with two and three facets to direct and vary the SHG signal, and find that the directionality of SHG can be varied and switched over 130 degrees. This ability is mainly facilitated by the single crystalline nature of the utilized mesowire, and is a clear advantage over polycrystalline nanostructures made of inorganic dielectric materials. In addition to directional SHG emission, we show selective nonlinear waveguiding capability of our geometry. Specifically we show how for the same pump excitation, the SHG signal is directional and confined at a location in the mesowire, whereas the two-photon excited fluorescence (TPEF) emission is isotropic and delocalized. This observation has direct consequence on spatial engineering of two different kinds of nonlinear emission on a single mesowire.\\

\section{Methods}
\subsection{Sample preparation} 
 1, 5 diaminoanthraquinone (DAAQ) mesowires were prepared through physical vapour transport method. The detail procedure to synthesize these molecular waveguide can be found in previous reports\cite{ravi-jopt}. In brief, DAAQ molecular powder (5mg, Sigma Aldrich, 85$\%$ pure) was dissolved in ethanol (60ml, Sigma Aldrich, 99$\%$ pure) in single neck round bottom (RB) flask. RB flask was then placed inside a heater cum rotor bath for depositing DAAQ molecular film around the RB wall. Continuous rotation at constant temperature (40$^\circ$C) ensures the uniformity of molecular thin film around wall of the flask. Afterwards, a cleaned glass coverslip was suspended inside the flask through a glass bar attached on the top of flask. Subsequently, the whole arrangement was then placed inside the silicon oil bath and increased the temperature of the system up to 160$^\circ$C-180$^\circ$C for different growth durations. As the temperature of the flask elevates, the molecules evaporate from the wall and start condensing on the glass coverslip. These condensed molecules act as preferred nuclei sites for new incoming molecules. Subsequently, these molecules get self-aggregated (J-aggregation takes place for our system\cite{ravi-jopt}, confirmed in our previous study) and resulted as extended one dimensional nanostructures (Fig. \ref{fig:2}(a)). The length and diameter of molecular waveguide can be easily controlled according to the growth duration and temperature. Deposition of wires was done on marked glass cover slip for the correlation between optical microscope image and field emission scanning electron microscope (FESEM) image of the wire to be studied. The optical absorption maxima of the grown mesowires was found to be around 520 nm (see Fig. \ref{fig:A1}, Appendix).  
\subsection{SHG microscopy coupled with Fourier plane imaging}

\begin{figure*}[h]
\includegraphics[scale=1]{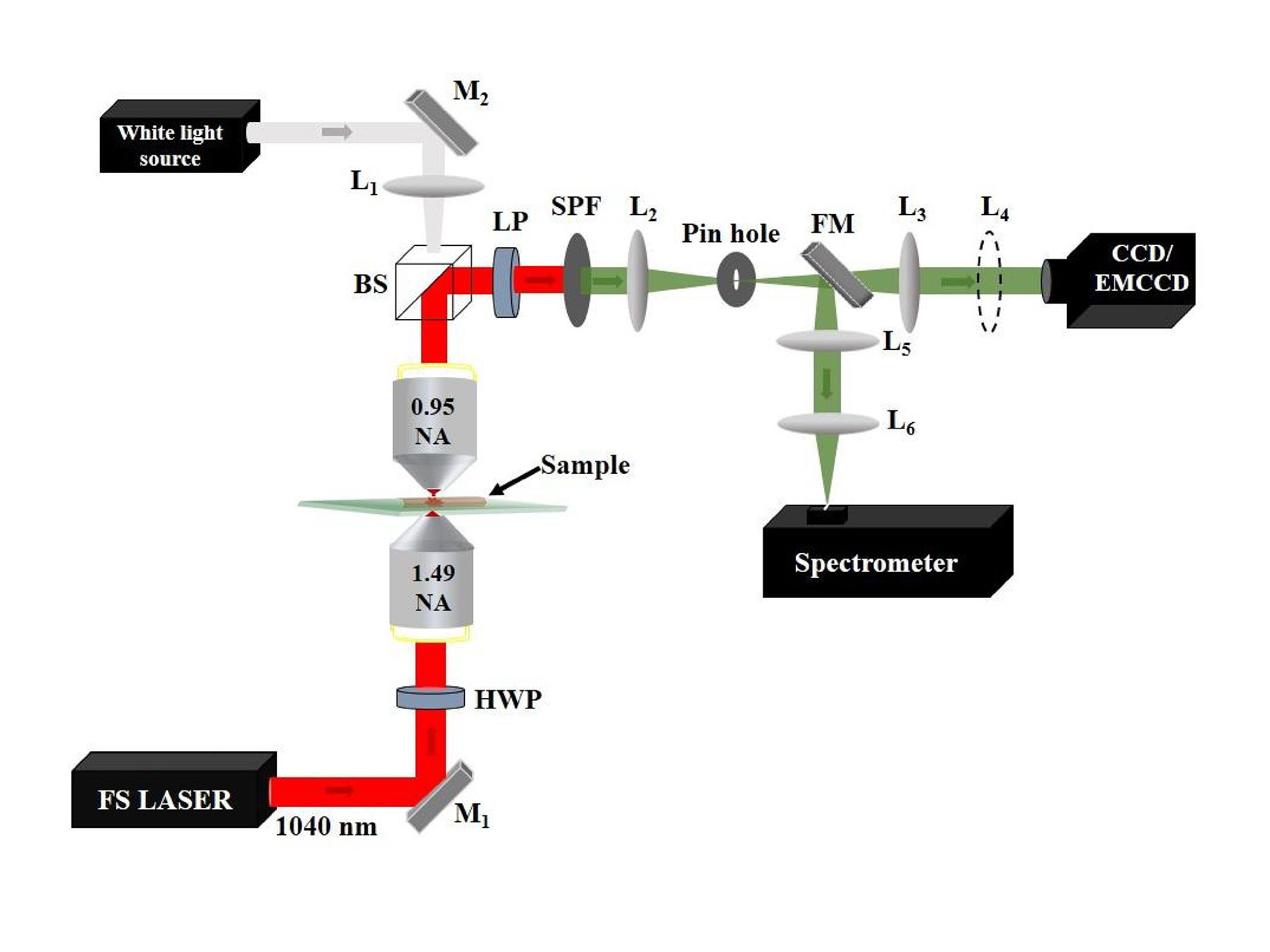} 
\centering
\caption{Schematic of Experimental setup for Fourier plane imaging and spectroscopy measurement of SHG. L1, L2, L3, L4, L5 and L6 are lenses where L2 is tube lens, L3 is Bertrand lens and lens L4 is used to form a real plane image. M1 and M2 are mirrors. FM is flip mirror. HWP, LP and SPF represent half wave plate, linear polarizer and short pass filter respectively. BS is beam splitter. SPF was used to block the pump. In case of Fourier plane imaging of SHG and TPEF, extra short pass filter and edge filter were used, respectively.}
\label{fig:1}
\end{figure*}

In a SHG microscope, the
captured SHG intensity can be expressed as\cite{brasaop}:
  \begin{equation}
I_{SHG} = |N \int_{V} \int_{\Omega}\int_{NA} E_{SHG}(r,\Omega, k)
f(\Omega) dr d\Omega dk |^2
  \end{equation}
  where $E_{SHG}$ is the electric field due of SHG signal,
$V$ is the focal volume from where the SHG is collected, $\Omega$ is
the orientation of the molecules within the focal volume, $k$ is
the wavevector of the SHG signal, $r$ is the spatial coordinate of
the nonlinear dipole, $f(\Omega)$  is the orientation factor, and
$NA$ is the numerical aperture of the collecting objective lens. In
our measurement, we collect the signal in the Fourier-plane of the
objective lens, such that we do not integrate the signal over the
whole NA, but retain the information of the in-plane $k$ vector
distribution. This $k$ vector distribution can be represented by two
polar angles $(\theta, \phi)$, where $\theta$ represents the radial
angle subtended by the numerical aperture of the lens, and $\phi$
represents the azimuthal angle. Figure \ref{fig:1} shows the optical microscope used to study the optical nonlinear response of the DAAQ mesowires. The lower objective lens ( high numerical objective lens, 1.49 NA, 100x) was used to excite the mesowire through the glass substrate using a Ti- sapphire laser (140 fs, 80 Hz, Chameleon from Coherent) operating at 1040 nm wavelength.

 The focused spot size was measured to be $\sim$ 670 nm. Measurements are done at 17 mW average power (at the entrance of microscope objective lens). We observed that exposing the wire with 58 mW or more power, burns the structure. Piezo Nanopositioning Stage (P-733.3CD, Physik Instrumente (PI) GmbH \& Co. KG) was used to scan the mesowire nm. Forward scattered signal was captured by the upper objective lens (0.95 NA, 100x). A pinhole was introduced in the real plane to spatially filter the signal from the point of excitation. For the Fourier plane imaging of the forward scattered optical nonlinear signals, back focal plane (BFP) of the upper objective lens was imaged to an electron-magnified charge-coupled device (iXon Ultra, Andor) using the tube lens (L2) and a Bertrand lens (L3). The pump (1040 nm) was blocked by a Short pass filter in the collection path. For Fourier plane imaging of SH and TPEF signals, we introduced additional short-pass and edge filters. The pump polarization was varied using a half wave plate (HWP) in the excitation path. A linear polarizer (LP) was used to study the output polarization dependence of the SH signal. For different excitation position, the BFP image was captured at a step of 100 nm.  A flip lens (L4) was used to capture the emission in real plane for each BFP image. For spectroscopic properties, the signal was sent to the transverse to its long axis over a diffraction limited excitation spot in the steps of 100 slit of the spectrometer using a flip mirror and was focused to spectrometer slit using lenses L5 and L6.
\subsection{Numerical calculations} 
 3D Finite Difference Time Domain (FDTD) simulations were performed at the pump wavelength (1040nm) by considering a triangular DAAQ waveguide resting on the glass substrate. Refractive indices (real and imaginary part) of DAAQ and glass were taken from \cite{ravi-jopt} and \cite{palik} respectively. The DAAQ wire was approximated as an irregular triangular waveguide of length 10 $\mu$m with one of the angles as 40$^\circ$ and base length as 1.3 $\mu$m (see Fig. \ref{fig:3}a). This was done to mimic the structure under study, as the wire clearly shows two facets in the top view (see SEM images in Fig. \ref{fig:3}b). The wire was illuminated using a Gaussian beam of wavelength 1040 nm through the glass substrate. The Gaussian beam was focussed at the glass-wire interface using a thin lens of NA 1.49.\\
The simulation area was meshed using the non-uniform conformal variant mesh of size 5nm to ensure accuracy and was terminated using Perfectly Matched Layers (PMLs) to reduce spurious reflections from boundaries. Near field electric field was calculated by positioning the Gaussian source at different positions along the axis perpendicular to the length of the wire.
\section{Results and discussion}
\subsection{DAAQ mesowires exhibit multiple facets} 
\begin{figure*} [h]
\includegraphics[scale=.75]{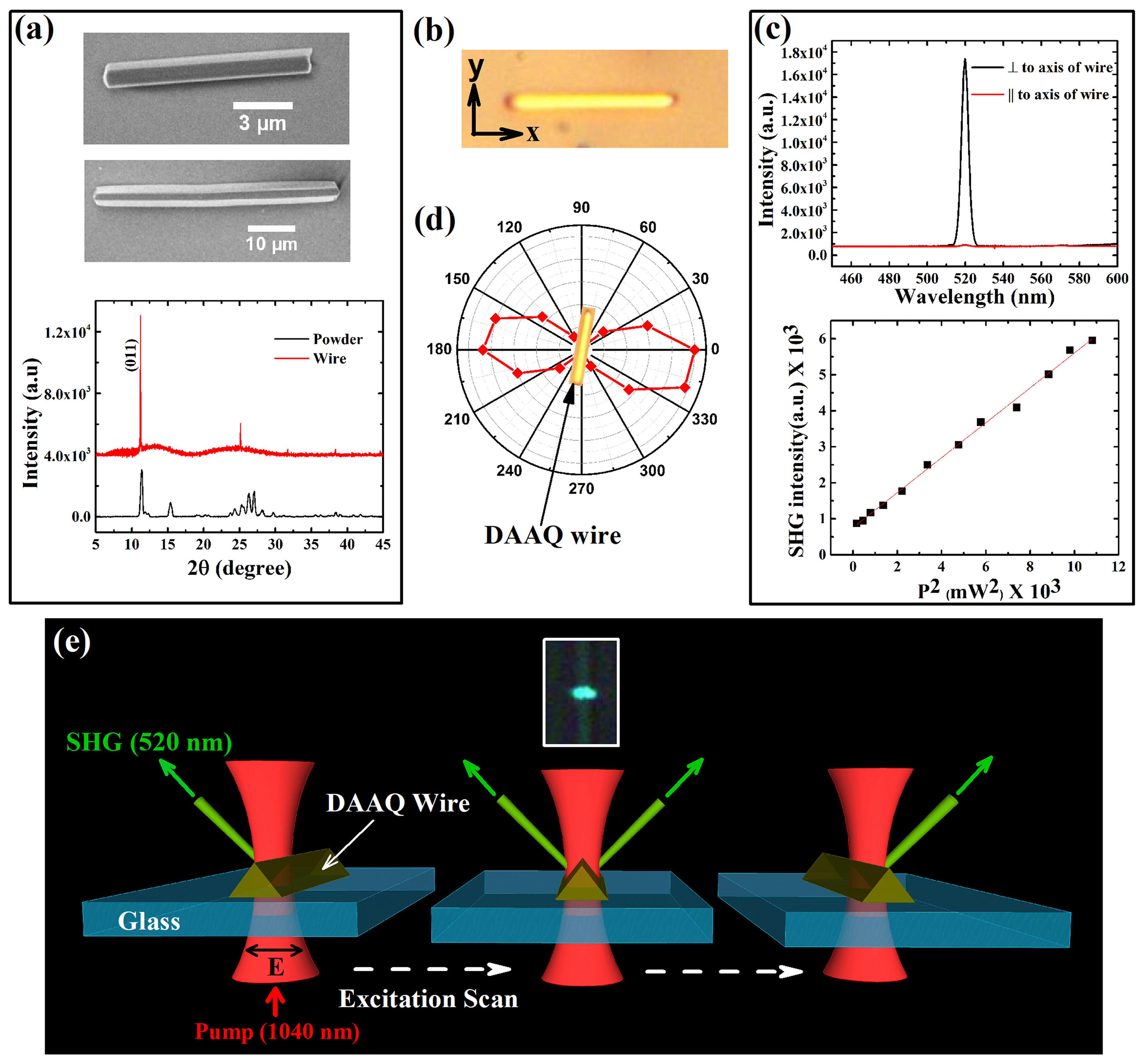}
\centering
\caption{(a) Representative FE-SEM images of faceted DAAQ mesowires. PXRD pattern for 1, 5 diaminoanthraquinone (DAAQ) mesowre on glass coverslip (red) and DAAQ powder(black). (b) Bright-field optical image of faceted DAAQ mesowire. The typical length is around 17 $\mu$m and width is 1.38 $\mu$m. (c) Forward-scattered SHG spectra from a DAAQ mesowire for excitation with pump polarization perpendicular and parallel to the long axis of wire (longitudinal). Pump excitation is 1040 nm (140 fs, 80 MHz). Power dependence of SHG signal: Peak intensity of the signal at 520 nm was measured as a function of pump power. (d) Variation of SHG as a function of angle of pump polarization. Maximum emission intensity is when excitation polarization is perpendicular to long axis of the mesowire. (e) Schematic of forward-scattered SHG wavevector variation as a function of position of 1040 nm focused pump beam across  the width of the DAAQ mesowire. The inset shows a real plane image of SHG emission at 520 nm from DAAQ mesowire when excitation beam is parked at the center of the wire.}
\label{fig:2}
\end{figure*}

Figure \ref{fig:2}a shows field-emission scanning electron microscopy (FE-SEM) images of typical DAAQ mesowires used in this study.
These mesowires are grown by vapour deposition method\cite{zhao1,rohit1,ravi-jopt} on a glass coverslip,
 and results in crystalline meso or nanowire \cite{zhaojacs}.
  The width of the wire can be controlled from meso- to nano-scale by varying the deposition time \cite{zhaojacs}.

  In order confirm crystallinity of the grown structure, we performed x-ray diffraction (XRD) on DAAQ mesowires deposited on a glass
  substrate (see XRD data in Fig. \ref{fig:2}a). The growth direction of mesowire was along [011] axis of the DAAQ crystal.

  The elongated crystals exhibit monoclinic lattice with parameters a = 3.78 {\AA}, b = 9.73 {\AA}, c = 15.01 {\AA}, $\beta$ = 82.4$^\circ$ \cite{zhaojacs} and centrosymmetric space group P2$_{1}$/C (no. 14)\cite{Yong}. From the top-view (Fig. 2a), most of the wires exhibit either two or three exposed facets. One of the main objectives of our work is to determine if the sub-wavelength facets of DAAQ mesowire can control the SHG radiation pattern.
\subsection{SHG characteristics of DAAQ mesowires} 
A typical optical bright field image of a faceted DAAQ mesowire
resting on a glass substrate is shown in Fig. \ref{fig:2}b. Since the width
of the wires are typically less than 2 microns, the facets cannot be
resolved in optical microscopy images (Fig. \ref{fig:2}b). First, we studied
the polarization resolved SHG spectra of DAAQ mesowires. By
employing a forward-scattering nonlinear optical microscope (see
methods section for details), we focused a 1040 nm femtosecond-laser
pump beam at the center of the mesowire, and captured the SHG signal
at 520 nm (after filtering out the TPEF signal) in the forward
scattering geometry. In Fig. \ref{fig:2}c, we resolve the SHG spectra with
respect to polarization of the excitation laser beam. The nonlinear
process is confirmed to be second order by measuring emission
intensity as a function of input power (see Fig. \ref{fig:2}c). When the polarization of pump beam is
perpendicular to the long axis of the mesowire, the SHG signal is
maximum, whereas for the parallel configuration, the signal is
minimum. In Fig \ref{fig:2}d, we show the pump-polarization dependent SHG
emission from the DAAQ mesowire. Going by the arguments of Brasselet
et al. \cite{brasaop}, two inferences that we can draw from this
observation are: a) the molecular dipoles of the mesowires are
aligned perpendicular to the long axis of the wire; b)
the distribution in orientation of the dipoles is narrow. We
anticipated no contribution towards emission from the bulk of
mesowire as DAAQ crystals used in this work have a center of symmetry.
One of the important aspect of DAAQ mesowire is that maximum SHG is emitted in forward direction. Figure \ref{fig:A1}b, Appendix shows SHG collected in forward (towards air) and backward (towards substrate) directions for a typical DAAQ mesowire. We have calculated the conversion efficiency (I$_{SHG}$ / I$_{Pump}$, where I: Intensity) of SHG in forward direction and found it to be of the order of 10$^{-9}$ for 20 mW pump power. Spectrometer is calibrated for pump power (1040 nm) by reducing the LASER power and sending it directly to the spectrometer. Efficiency is calculated by taking the ratio of the counts of SHG in the spectrometer and counts corresponding to the pump power (20 mW) measured at the entrance of objective lens. We have corrected for the response of the spectrometer at SHG and pump wavelengths. It should be noted that calculations for efficiency do not correct for losses due to optical components. 
\begin{figure*} [h]
\includegraphics[scale=1]{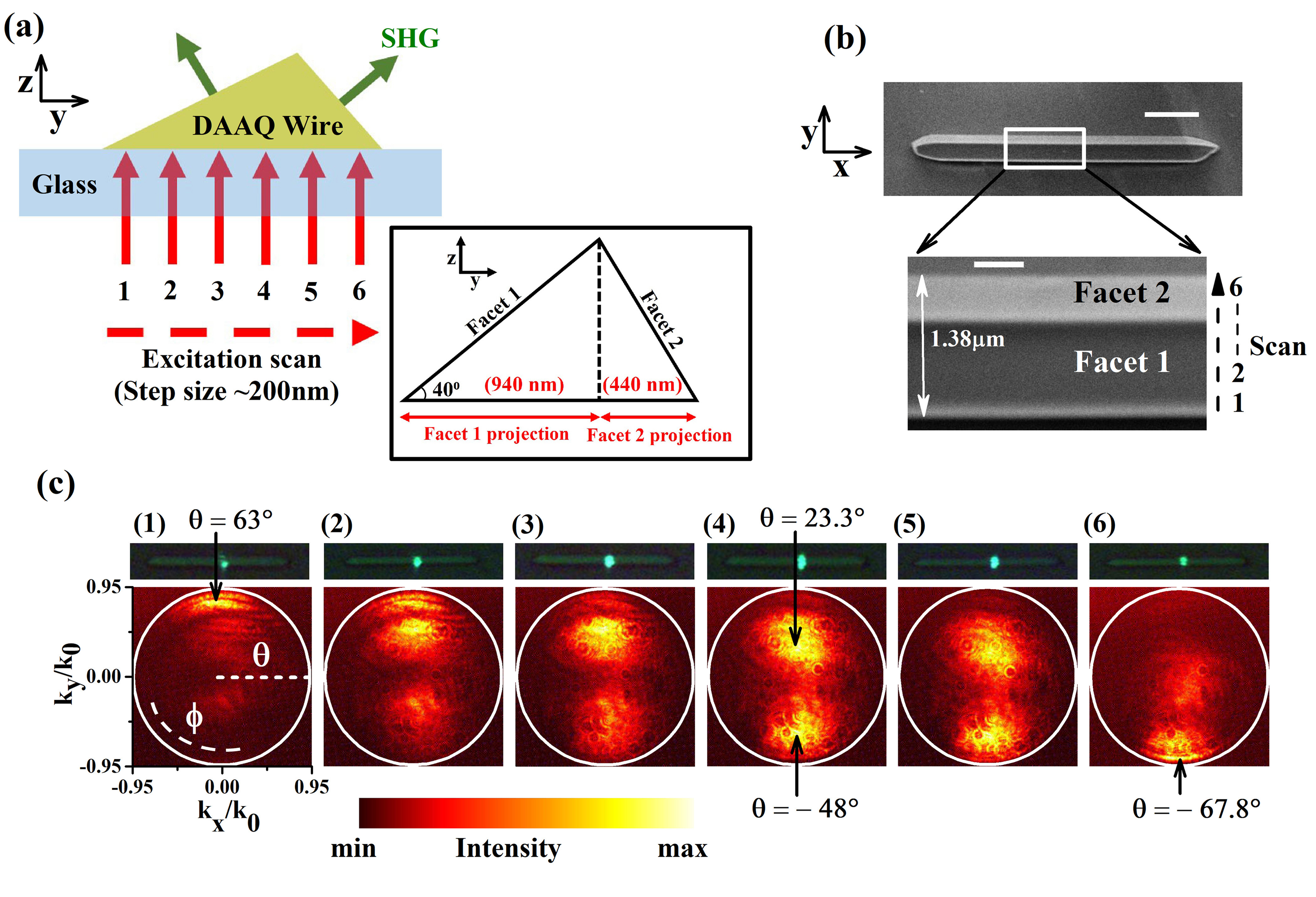}
\centering
\caption{(a) Schematic of the experiment in which the focused excitation position is varied across the width of a DAAQ mesowire with two facets facing the air superstrate.  Tip of the red arrows 1 to 6 indicate the excitation spot. Green arrows indicate SH emission from facets of the wire. Inset shows that projections of facets on base have different widths which further indicate that one of the angles is less than 60$^\circ$. This angle is assumed to be 40$^\circ$ for numerical calculations.  (b) FESEM images of DAAQ mesowire of $\sim$16.5 $\mu$m length at two different magnifications. Scale bars are of length 3 $\mu$m (full-scale image) and 500 nm (zoomed in image)  (c) SHG emission (real plane images) and corresponding wave vector distributions (Fourier-plane images) for excitation positions 1 to 6 shown in Figs. (a) and (b). $\theta$ is half angle of captured cone and has maximum value of 71.8$^\circ$ defined by numerical aperture (0.95) of objective lens. $\phi$ is azimuthal angle. Wavevector axes of Fourier-plane image are defined as k$_{x}$/k$_{0}$=sin$\theta$ cos$\phi$ and k$_{y}$/k$_{0}$=sin$\theta$ sin$\phi$.}
\label{fig:3}
\end{figure*}
\subsection{Hypothesis and scheme of the experiment} 
Given that the mesowire have subwavelength facets, how do they influence the directionality of the forward SHG emission? This is the central question we ask, and Fig. \ref{fig:2}e shows the schematic of our experiments to address this question. Our hypothesis is that when we precisely scan the pump excitation (in steps of around 100 nm) across the width of the mesowire, we should be able to see a drastic change in the forward SHG radiation pattern. This expectation is due to the presence of the facets on the crystal, which should influence the emission directionality. Importantly, we are interested to know if we can switch the directionality of the SHG emission when the pump beam was moved from one edge of the mesowire to another (see schematic in Fig. \ref{fig:2}e). In the inset of Fig. \ref{fig:2}e, we show a typical real colour image of the SHG emission at 520 nm when the pump beam is focussed at the center of the mesowire. When the beam is laterally scanned across the width of the mesowire, we do observe spatial modulation of SHG emission (see Visualization 1). Having observed this modulation, our next goal is to quantify the angular intensity profile and correlate it to the morphology of the mesowire.
\begin{figure}
\includegraphics[scale=.8]{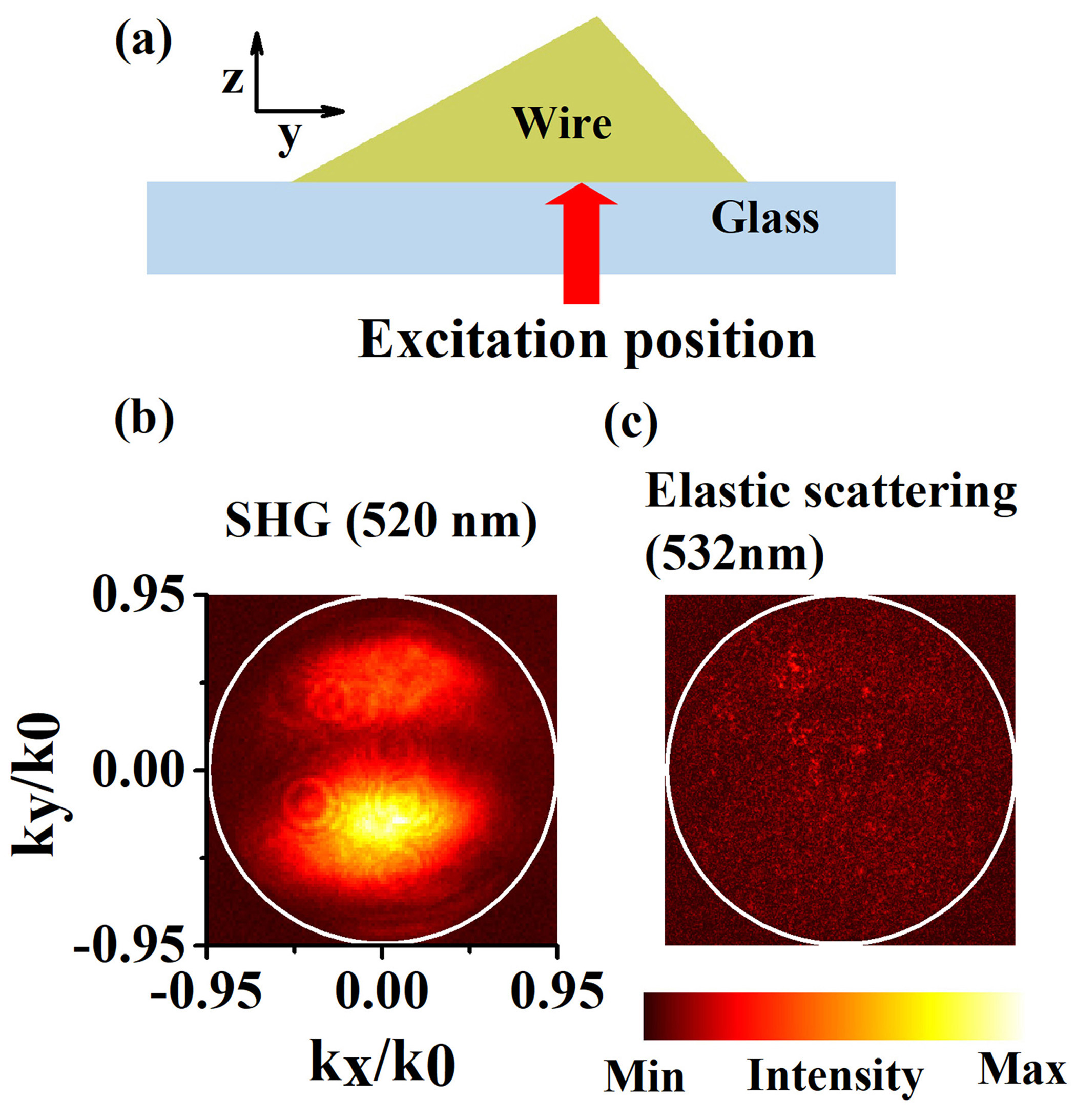}
\centering
\caption{Comparison of angular intensity distribution of SHG scattering (520 nm) and elastic scattering (532 nm) from the same DAAQ mesowire. (a) Schematic showing position of the pump beam ( 1040 nm for SHG scattering and 532 nm for elastic scattering) at the center of the mesowire. Fourier plane images captured for (b) SHG scattering and (c) elastic scattering reveals the difference in angular intensity distribution.}
\label{fig:4}
\end{figure}
\subsection{Variation of SHG wavevector from a DAAQ mesowire with two-facets}
First, we begin by discussing the effect of varying the excitation position on the angular emission of the second-harmonic generated at the DAAQ mesowire. Figure \ref{fig:3}a shows the schematic of the experiment performed, in which the SHG signal is captured by positioning the pump-beam at various locations across the width of the DAAQ mesowire. Positions 1 to 6 indicate the different locations of the focused pump beam (step size$\sim$200 nm). The SHG emission pattern is angularly resolved by a transmission mode Fourier microscope (see methods for the details on instrumentation). The DAAQ mesowire that we used for the experiment is visualized by FE-SEM imaging at two different magnifications (Fig. \ref{fig:3}b).\\
In Fig. \ref{fig:3}c (top) we show a series of real-plane SHG images recorded from the mesowire for the six different locations of the laser spot. The numbers indicate various positions ($\sim$ 200 nm step size) of the excitation beam that result in SHG scattering. Fig. \ref{fig:3}d (bottom) represents the corresponding Fourier-plane image showing the angular distribution of SHG signal.
  Following are inferences that can be drawn from the experiment:
  \begin{itemize}
    \item By changing the position of excitation, the directionality of the SHG scattering wavevector can be varied. Fig. \ref{fig:3}d shows an angle of maximum intensity at $\theta$ = 63$^\circ$ for the excitation focus at one of the lateral edges of wire (position 1, Fig. \ref{fig:3}b). When the focus is moved to the other end of the lateral edge of the mesowire (position 6, Fig. \ref{fig:3}a) the angle of maximum intensity is at $\theta$ = -67.8$^\circ$. This indicates a change of 130.8$^\circ$ in the SHG emission angle. We note that the maximum angle (71.8$^\circ$) captured in our experiment is limited by the numerical aperture of the collection objective (0.95 in this case). This means there can be SHG emission beyond 71.8$^\circ$ which is not recorded in our measurements. At position 4, when the excitation is at the center of the mesowire, maximum SHG emission occurs at angles $\theta$ = 23$^\circ$ and $\theta$ = -48$^\circ$.

     \item The directionality of the SHG scattering depends upon the inclination of the facets with respect to the excitation beam.

\item For certain excitation positions (number 2 in Fig. \ref{fig:3}c), we observed interference fringes in the Fourier-plane image. This may be due to coherent superposition of the SHG waves emanating from dipoles at two different positions on the mesowire.
\item Added to this, the grown mesowires have strong absorption at the wavelength of SHG emission (520 nm in this case). As a result, the
angular distribution of the surface SHG scattered light (520 nm) drastically differs from elastic scattered light (532 nm) (Fig. \ref{fig:4}). Interestingly, the wire behaves as an opaque object (width $\sim$ 3$\lambda$) at 532 nm excitation close to the absorption maximum of DAAQ mesowire. Fig. \ref{fig:4}b shows minimally transmitted light (background speckles) through the wire. In contrast, the SHG scattered light at 520 nm shows facet-sensitive directional emission (Fig. \ref{fig:4}a). It should be noted that elastic scattering at 532 nm wavelength was performed using a line filter (for 532 nm wavelength) in the collection path of the microscope to filter out any secondary emission.
\end{itemize}

Also, we performed angle resolved forward SHG experiment on a DAAQ mesowire with three facets, and found systematic variation of SHG as we move the excitation beam across the width of the mesowire (Data not shown).
\subsection{Parameters influencing the SHG wavevector}
Having experimentally observed variation in the SHG radiation pattern as a function of excitation location, we are interested in understanding the relevant parameters that can be connected to our observations. From a theoretical basis on nonlinear microscopy, the SHG radiation patterns from a molecular assembly can be studied from at least two approaches. One is the phased dipole array approach \cite{mertzjosa, Moreaux, moreaux2,mertz20} and the other is the Green$'$s tensor approach \cite{Yew}. We adapt the former approach to understand basic elements of our results. In a SHG microscope, the far-field power per differential solid angle is given by:\cite{mertzjosa}
\begin{equation}
P_{2\omega}(\theta, \varphi)  = \frac{1}{2}  n_{2\omega}
\epsilon_{0}  c  r^2   |E_{2\omega}(\theta, \varphi)|^2;
\end{equation}
where $n_{2\omega}$ is the refractive index of the mesowire at the SHG frequency,  $r$ is distance between the nonlinear dipole and the detector, $E_{2\omega}(\theta, \varphi)$ is the electric field of the SHG. This field can be further expressed as:\cite{mertzjosa}
 \begin{equation}
 E_{2\omega}(\theta, \varphi) = E_{2\omega}^{(0)}(\theta, \varphi)\cdot N \cdot A(\theta, \varphi).
\end{equation}
  The $E_{2\omega}^{(0)}(\theta, \varphi)$ represents the dipolar electric field emission at the focal center of the illumination, $N$ represents the total number of molecules in the focal volume contributing to the SHG signal i.e. molecules present on the facet of the wire, and $A(\theta, \varphi)$ is a scalar function representing the angular modulation, and is dependent on the waist of the excitation beam and the Gouy phase shift at the point of excitation \cite{mertzjosa}. Equation (3) suggests that far field distribution depends on the dipole orientation. Also when we move our focused excitation along the base of the mesowire, both $N$ and $A$ (dependent on the geometry of the structure) vary, which leads to variation in the detected SHG radiation pattern. We note that apart from the general parameters discussed above, the cross-section geometry also plays a vital role in our observations. Especially, the orientation of the facets with respect to the excitation beam will influence the SHG pattern, which is discussed below.
\subsection{Near-field excitation profile highlights the role of the facets}
\begin{figure*} [h]
\includegraphics[scale=1]{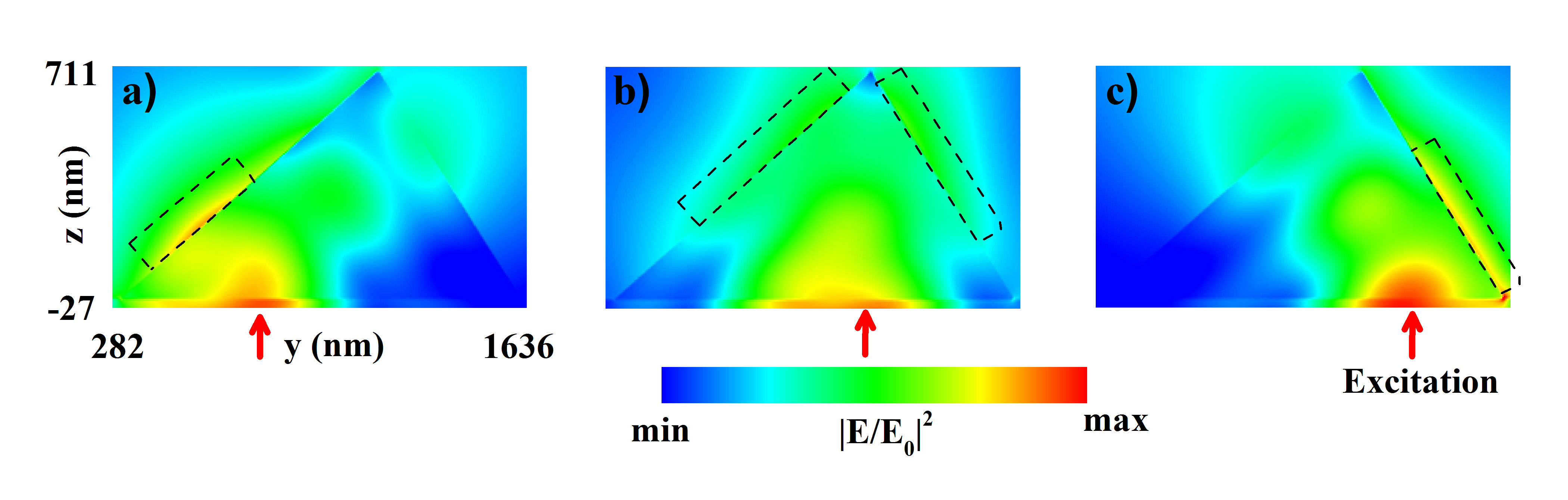}
\centering
\caption{FDTD simulations showing electric near-field distribution at excitation wavelength (1040nm) for various positions (a-c) along the cross-section of a two-facet mesowire geometry. The arrows indicate the position of Gaussian beam excitation. The dotted boxes indicate the region of maximum intensity at the facets of the mesowire}
\label{fig:5}
\end{figure*}
  To further understand the change in the wave vector distribution of SH emission as a function of excitation scan, we simulate the pump field distribution for different lateral positions of excitation. Full-wave Finite Difference Time Domain (FDTD) calculations are performed to visualize pump-field distribution at the upper two facets of DAAQ mesowire (see methods section for details on simulation). The real and imaginary parts of the refractive index of the mesowire used in the simulations are extracted from past measurements \cite{ravi-jopt}. The maximum value of the near-field distribution of the pump-field (1040 nm) changes as we change the position of excitations in the geometry (see dotted boxes in Fig. \ref{fig:5}a to c). Since the orientation of the facets changes as we proceed from right to left in the geometry (Fig. \ref{fig:5}), the location of the nonlinear dipole excitation and emission will vary accordingly. What essentially is transferred to far-field is the emission from the contributing nonlinear dipole at the facets. The directionality of the light in far-field now depends on which facet dominantly contributes towards the nonlinear emission. A conclusion that we can draw from this result is that by changing the excitation position one can selectively excite specific dipoles in the geometry, which further contributes towards directional SHG signal emanating from the object. Although FESEM images (Fig. \ref{fig:2}a and \ref{fig:3}b) show a smooth surface of DAAQ mesowires, there may be some fine structural details that are not detectable at the given resolution of electron microscope. These structures can potentially influence electric near-field of pump LASER and may further affect SHG from facets of the mesowire. It is to be noted that for the simulations, we have not considered such fine structural details of the wire.
     The experiments and simulations indicate that by varying the position of the excitation-pump in a nonlinear microscope, one can alter the transmitted SHG wavevector. Such capability to precisely tune the directionality of SHG has direct implications on nonlinear optical antennas, where the directionality of the nonlinear scattering wavevector can be actively tuned by external optical parameters without having to alter the geometry, composition or shape of the material. We emphasize that such spatial tunability is mainly due to the fact that self-assembled organic mesowires facilitate inclined crystallographic facets that can be harnessed to control directionality of nonlinear optical antennas.
    
\subsection{A prospect for spatial engineering of nonlinear emission: SHG and TPEF from DAAQ mesowire}
\begin{figure*} [h]
\includegraphics[scale=1]{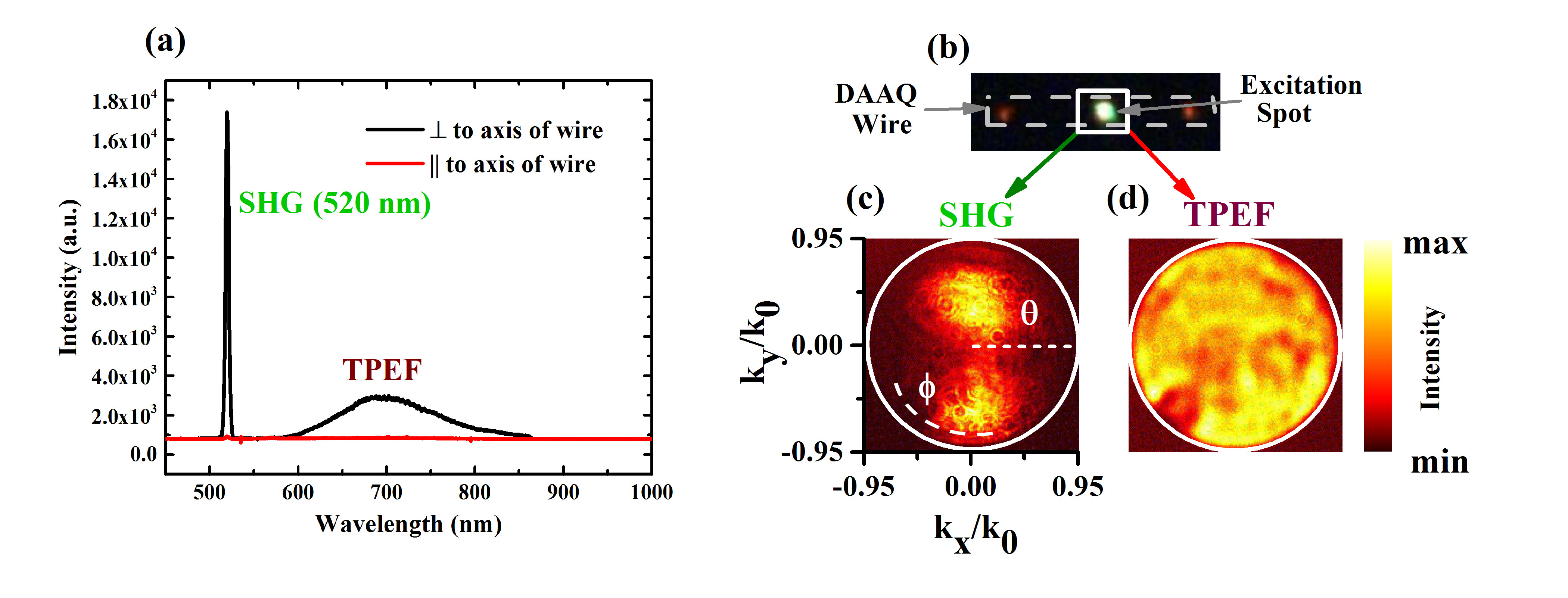}
\centering
\caption{(a) SHG and TPEF spectra from a DAAQ mesowire for pump polarizations parallel and perpendicular to the long axis of the wire. (b) Nonlinear optical image of a DAAQ mesowire excited at the center. Note there is faint TPEF emission from distal ends of the mesowire, which indicates waveguiding property of the mesowire. Fourier-plane images of the (c) SHG and (d) TPEF emission from the center of the wire indicates directional and isotropic distribution of light, respectively. These images were collected by spatial and spectral filtering techniques.}
\label{fig:6}
\end{figure*}
There are two aspects of organic molecular wires which we wish to emphasize. First is that they facilitate not only SHG but also two-photon-excited fluorescence (TPEF) \cite{brasaop}. Second is that certain organic mesowires can function as an optical waveguide of fluorescent signal \cite{taka,ravi-jopt}. With this hindsight, we are interested to compare the directionality of SHG and TPEF from the same excitation on DAAQ mesowire, and observe waveguiding effect, if any. Figure \ref{fig:6}a shows the optical spectra of the TPEF and SHG collected from the center of a DAAQ mesowire (Fig. \ref{fig:6}b) with excitation parallel and perpendicular to the mesowire. We observed that the SHG emission is typically more intense compared to TPEF for both the cases. Both emission processes reach their maximum intensity when pump polarization is oriented perpendicular to long axis of the mesowire. This again confirms the orientation of nonlinear dipole to be perpendicular to the long axis of the wire.  Next, we probed the directionality of the TPEF compared to SHG signal. To this end, we captured angle-resolved nonlinear scattering by selective spectral and spatial filtering from the region of interest and imaging it on the Fourier-plane of the collection objective lens (see method section for details). In Fig. \ref{fig:6}c and d we compare the angle resolved SHG and TPEF, respectively. It is clear that SHG scattering is directional compared to isotropic emission due to TPEF. This directional emission of SHG is because of the fact that SHG is a coherent emission \cite{brasaop} process with a specific phase defined between the nonlinear dipolar emitters, whereas TPEF is an incoherent process. Another observation is that the TPEF signal propagates along the length of the mesowire and outcouples from ends of the wire, whereas the SHG is confined mainly to the location of excitation (see the center and distal ends of the mesowire in Fig. \ref{fig:6}b). This observation of localizing the coherent signal and delocalizing the incoherent nonlinear optical emission for a common pump-excitation can be harnessed in signal processing, where the coherent source is to be out-coupled in a specific direction and incoherent signal is to be guided along the circuit.
\section{Conclusion}
In summary, controlling the directionality of nonlinear optical signal from sub-wavelength structures have direct ramification on design and development of active optical antennas and nonlinear optical circuits. Our study shows, how the crystallographic facets of an organic mesowire can be utilized to control the second harmonic generation emission pattern at sub-wavelength scales. We have demonstrated this principle on a two-facet and three-facet mesowire crystals, and have emphasized the role of selective excitation of spatially distributed dipoles as an important criterion for our observations. Numerical simulations of the prototypical geometry show spatial dependence of the near-field excitation, which further corroborate the assumptions that nonlinear dipoles distributed on the facets of the extended mesowire play a vital role in controlling the SHG directionality. Given that organic molecular mesowires can be prepared and deposited on various substrates, our observation can be adapted to various platforms including flexible plasmonic metasurfaces. An attractive aspect of organic mesostrutures is that they can be easily integrated on various optoelectronic circuits, which means the proposed optical antennas can be potentially operated under electric bias. Such biasing mechanisms can be utilized not only as active optical antennas, but also as coherent light emitting devices facilitated by Frenkel exciton-polaritons\cite{ravi-jopt} in such organic nanostructures. Furthermore, our results show how SHG and TPEF can be spatially and angularly discriminated on the mesowire by utilizing the waveguiding capability of the quasi-one dimensional structure. Such spatial and angular discrimination of two different nonlinear optical emission processes by the same excitation can provide interesting opportunities in nonlinear optical circuitry. An interesting prospect of our work is to couple organic mesowires to optical microcavities, where mode coupling between the wire and the cavity can lead to nonlinear optical interaction which can be reversibly switched from weak to strong coupling regimes.
\section{Funding Information}
This research work was partially funded by DST-Nano mission Grant (SR/NM/NS-1141/2012(G)), Center For Energy Science (SR/NM/TP-13/2016), AFRL grant - GRANT12146178 and the financial support of  Indo-French grant (IFCPAR), project $\#$5504-3.

\appendix
\section{Appendix}
This section provides information about characterization of DAAQ mesowires. 
\renewcommand{\thefigure}{A\arabic{figure}}
\setcounter{figure}{0}
\begin{figure} [h]
\includegraphics[scale=1]{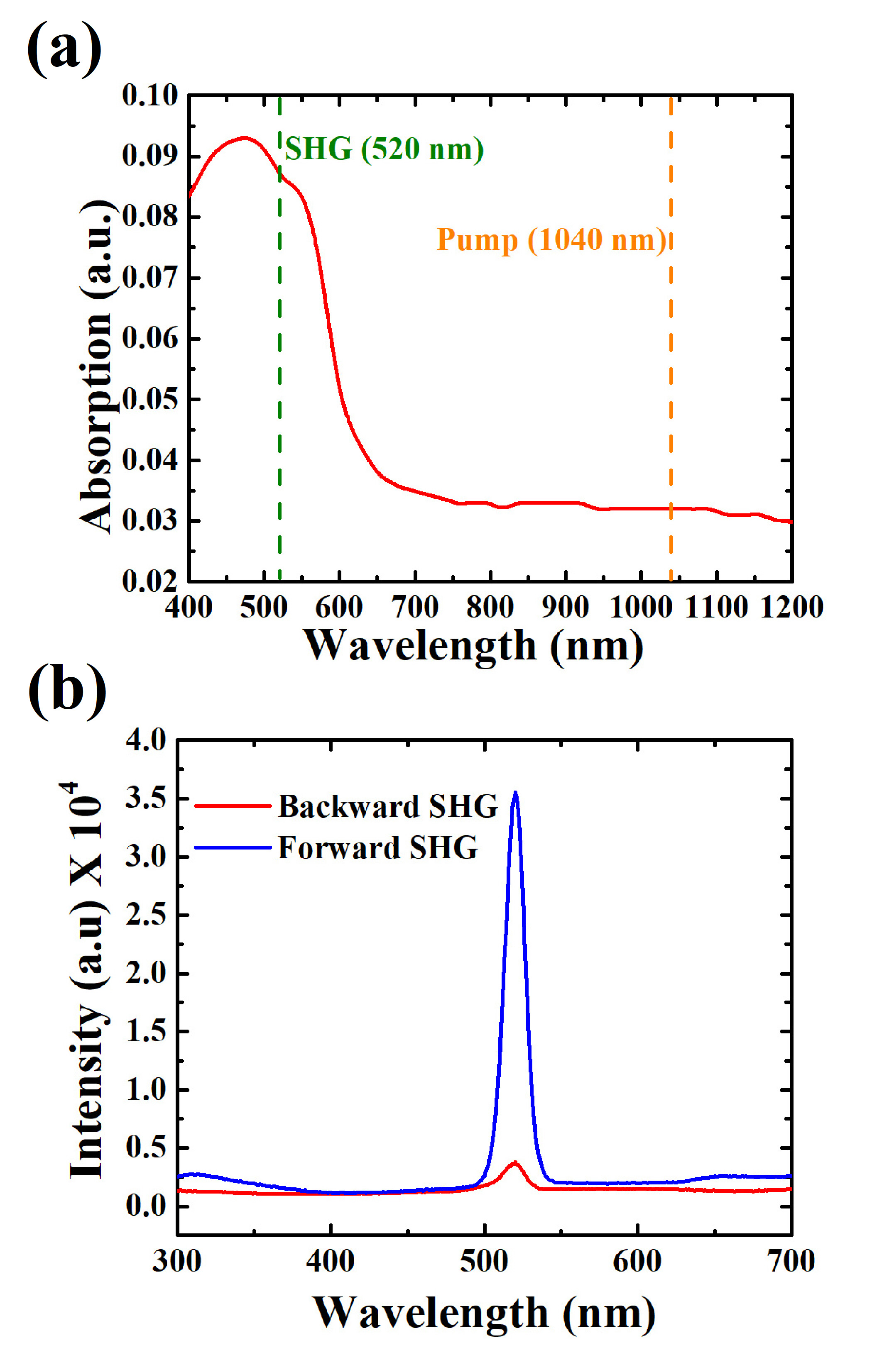} 
\centering
\caption{ (a) Absorption Spectra of DAAQ mesowires placed on the glass substrate (red curve). (b) SHG from DAAQ mesowire collected in forward direction (blue curve) and backward direction (red curve). 
}
\label{fig:A1}
\end{figure}

\bibliographystyle{unsrt}

\bibliography{sample1}

\end{document}